\documentstyle[12pt]{article}

\begin{document}
\begin{center}
{\Large \bf Fractal universe }
\bigskip

{\large D.L.~Khokhlov}
\smallskip

{\it Sumy State University, R.-Korsakov St. 2, \\
Sumy 244007, Ukraine\\
E-mail: others@monolog.sumy.ua}
\end{center}

\begin{abstract}
The model of the universe
is considered in which background of the universe
is not defined by the matter but is a priori specified as a
homogenous and isotropic flat space.
The scale factor of the universe follows
the linear law. The scale of mass changes proportional to
the scale factor. This leads to that
the universe has the fractal structure with a power index of 2.
\end{abstract}

Pietronero and collaborators~\cite{Pi}, \cite{Sy}, \cite{Ba}
performed the analysis of galaxy and cluster correlations
with the use of the following catalogues: CfA,
Perseus-Pisces, SSRS, IRAS, LEDA, Las Campanas and ESP for galaxies
and Abell and ACO for clusters.
They determined the quantity
\begin{equation}
N(<R) = \int_0^R dr \, \sum_{i} \delta(r-r_i) \; \propto \; R^D
.\label{eq:monop}
\end{equation}
The analysis showed that all the
available data are consistent with each other and reveal fractal correlations
(with dimension $D \simeq 2$) up to the deepest scales probed until now ($1000
h^{-1}{\ \rm Mpc}$) and even more as indicated from the new
interpretation of the number counts.

In~\cite{Kh}, the model of the universe with the linear law of evolution
was considered in order to resolve the flatness and horizon problems.
The main idea is that the background of the universe is not
defined by the matter but is a~priori specified.
Homogenous and isotropic flat space with the spatial metric $dl$
and absolute time $t$ is taken as a background of the universe
\begin{equation}
dl^2=a(t)^2(dx^2+dy^2+dz^2), \quad t,
\label{eq:met}
\end{equation}
with the scale factor of the universe is a linear function of time
\begin{equation}
a=ct.\label{eq:g1}
\end{equation}

The total mass of the universe relative to the background space
includes the mass of the matter and the energy of its gravity.
Let the total mass of the universe be equal to zero,
then
the mass of the matter is equal to the energy of its gravity
\begin{equation}
c^2={GM\over{a}}.\label{eq:o}
\end{equation}
That is the mass of the matter is a linear function of
the scale factor
\begin{equation}
M={c^2a\over{G}}.\label{eq:p}
\end{equation}

Take the volume of the radius $R_{1}$. Define the mass density of
the volume as
\begin{equation}
\rho_{1}=\frac{m_{1}}{R_{1}^3}.\label{eq:rh1}
\end{equation}
Take the volume of the radius $R_{2}$. Define the mass density of
the volume as
\begin{equation}
\rho_{2}=\frac{m_{2}}{R_{2}^3}.\label{eq:rh2}
\end{equation}
In view of Eq.~(\ref{eq:p}), to compare these two densities
one should take into account the change of the scale of mass.
Then Eq.~(\ref{eq:rh2}) should be rewritten as
\begin{equation}
\rho_{2}=\frac{m_{2}}{R_{1}R_{2}^2}.\label{eq:rh21}
\end{equation}
If to think that $\rho_{1}=\rho_{2}$, then the volume within
radius $R_{2}$ is given by
\begin{equation}
V(<R_2)=\frac{m_{2}}{\rho_{2}}=R_{1}R_{2}^2.\label{eq:v2}
\end{equation}
It is natural to take the volume of the universe as a standard one.
Then the volume within radius $R$ is given by
\begin{equation}
V(<R)=aR^2.\label{eq:vR}
\end{equation}

Let us define the number particle density in the universe as
\begin{equation}
n=\frac{N(<a)}{a^3}.\label{eq:n}
\end{equation}
Then the number of particles within radius $R$
is given by
\begin{equation}
N(<R)=nV(<R)=naR^2=N(<a)\frac{R^2}{a^2}.\label{eq:NR}
\end{equation}
So the universe has the fractal structure with the power index $D=2$.

\end{document}